%% file: Paper1.9.tex
\documentclass[useAMS,usenatbib,fleqn]{mn2e}

\usepackage{times}
\usepackage{hyperref}
\usepackage{tikz}
\usetikzlibrary{shapes,arrows}
\usepackage{graphicx}
\usepackage{amssymb}
\usepackage{amsmath}
\usepackage{amsfonts} 

\hypersetup{pdftitle={The VIPERS project: 
classifying VIMOS low-resolution spectra through Principal Component Analysis},
  pdfauthor={Alida Marchetti},
  hyperfootnotes=true,
  colorlinks=true,
  pdfborder={0 0 0},
  citecolor=black,
  urlcolor=blue,
  filecolor=blue,
  linkcolor=black,
}

\newcommand{\simlt}{\,\hbox{\lower0.6ex\hbox{$\sim$}\llap{\raise0.6ex\hbox{$<$}}}\,}

\title[VIPERS: 
classifying VIMOS low-resolution spectra through PCA]{
The VIMOS Public Extragalactic Redshift Survey (VIPERS): 
spectral classification through Principal Component
Analysis\thanks{based on observations collected at the European Southern Observatory, Cerro Paranal, Chile, using the Very Large
    Telescope under PID 182.A-0886}
}

\include{authors_gigi_work}

\begin{document}

\pagerange{\pageref{firstpage}--\pageref{lastpage}} \pubyear{2012}

\maketitle
\label{firstpage}

\clearpage

\begin{abstract}
We develop a Principal Component Analysis aimed at classifying a
sub-set of 27,350 spectra of galaxies in the range $0.4<z<1.0$
collected by the VIMOS Public Extragalactic Redshift Survey (VIPERS).
We apply an iterative algorithm to simultaneously repair parts of
spectra affected by noise and/or sky residuals, and reconstruct gaps
due to rest-frame transformation, and obtain a set of orthogonal
spectral templates that span the diversity of galaxy types.  By taking
the three most significant components, we find that we can describe
the whole sample without contamination from noise.  We produce a
catalogue of eigen-coefficients and template spectra that will be
part of future VIPERS data releases.  Our templates effectively
condense the spectral information into two coefficients that can be
related to the age and star formation rate of the galaxies.  We
examine the spectrophotometric types in this space and
identify early, intermediate, late and starburst galaxies. 
\end{abstract}

\section{Introduction}

Galaxies can be largely divided into two
classes: early type galaxies, characterized mainly by old, passively
evolving stellar populations, and late type galaxies that show
evidence for recent star formation.  This dichotomy is displayed in
the local Universe in the morphology of galaxies \citep{Sandage75}, as well
as in their colours \citep{vauc62}, spectral
characteristics \citep{MM57, Mad02} and clustering properties
\citep{Davis76, Giovanelli86, Gigi97, Nor2002, Phleps2006, Coil2006, Meneux2008, Meneux2009,
  Zehavi2011}.  This is already present at high redshifts
\citep{Brown2003, Daddi2003, Coil2008, Abbas2010,
  delatorre2011,Coupon2012} and provides 
fundamental constraints on galaxy formation and evolution models.  The
distribution of galaxy colours is observed to be bimodal, with two
distinct peaks in the red and in the blue \citep{Strateva01,
Bell2004, Baldry2004, Weiner2005, Faber2007, Franzetti2007}. Between
these classes lie galaxies with intermediate 
colours in the \emph{green valley}.  These share the characteristics
pertaining to both red and blue classes and are thought to be caught
during the transition from a period of active star formation to
quiessence \citep{Bell2004, Baldry2004, Faber2007, Brammer2009}.

Spectroscopy provides a deeper insight into the physics of galaxies,
with respect to average colours determined from broad band photometry.
For example, selecting red galaxies solely on broad-band colours does not result in a sample
of dead, passive early-type objects but also contains a non-negligible fraction
of star forming galaxies and/or dusty starbursts \citep{Cimatti2002, Gavazzi2003, Franzetti2007, Graves2007}.
Conversely, the high information content of the data set makes it
difficult in general to compress and classify all the information
contained in a galaxy spectrum in a compact and efficient way.
Statistical methods have been successfully used to reduce such
complexity by identifying specific features, such as
emission line intensities or continuum break strengths (e.g.
\citealt{Mad2003}).  An important method to identify the essential
information from complex multi-dimensional datasets is represented
by Principal Component Analysis (PCA).  Each galaxy spectrum is
linearly decomposed into a set of representative templates. 
The PCA naturally determines the minimum number of templates required
to describe 
the sample given the noise properties of the spectra.  These templates
show the features of the spectra that have the most discriminating
power (the Principal Components). For astronomical 
spectra, the Principal Components have been shown to characterize well
the spectral slope, and the presence of strong emission lines,
allowing the sample to be divided into classes.  Often these classes
correspond to physical characteristics of the galaxy and can
distinguish star-forming, post-starburst and passive galaxies
\citep{Connolly95,Ferreras2006,Rogers2007,Rogers2010}.

The PCA has been applied to classify galaxies from the Sloan Digital
Sky Survey (SDSS, \citealt{York2000}) \citep{Yip2004, Dobos2012}. The
effectiveness of the method was confirmed well before in the
separation of broad absorption line QSOs from a full QSO sample
\citep{Francis93}, the classification of spectral energy distributions
for stars \citep{SGG98}, or the classification of other galaxy spectra
\citep{Folkes96, Sodre97, Bromley98, Galaz98,  Ronen99}. 

In particular, Folkes, Lahav and Maddox in 1996 investigated low
signal-to-noise spectra with the PCA technique and reconstructed the
underlying physical information using only 3 components. Combining the
results of the PCA with a neural network approach they successfully
classified a group of simulated spectra into different morphological
classes.

Furthermore, Connolly and Szalay in 1995 carried out a classification
of ten template galaxy energy distributions in terms of an orthogonal
basis, to estimate the number of significant spectral components that
comprise a particular galaxy type, finding a correlation between their
spectral classification and those determined from published
morphological classifications. 

The application of classification methods to observed galaxy spectra
presents some challenges.  Spectra can be affected by spurious noise
features, as positive or negative line residuals due to poor sky
subtracion. This is the case of VIMOS spectra prior to August 2010,
due to the fringes produced by interference of bright sky
lines with the CCD surface. Other features can be the result of
zero-order images of bright objects from adjacent spectra.  All these
features may have been corrected to some extent in the processed
spectra, or be still present in the spectra.  The many disguises these
artefacts can take make it difficult to accurately classify spectral features.   
We will show that through
the application of PCA we can accomplish the task of cleaning the
spectra of noise artefacts while simultaneously obtaining a
classification by means of a handful of parameters.  

\begin{figure}
\includegraphics[scale=1.2]{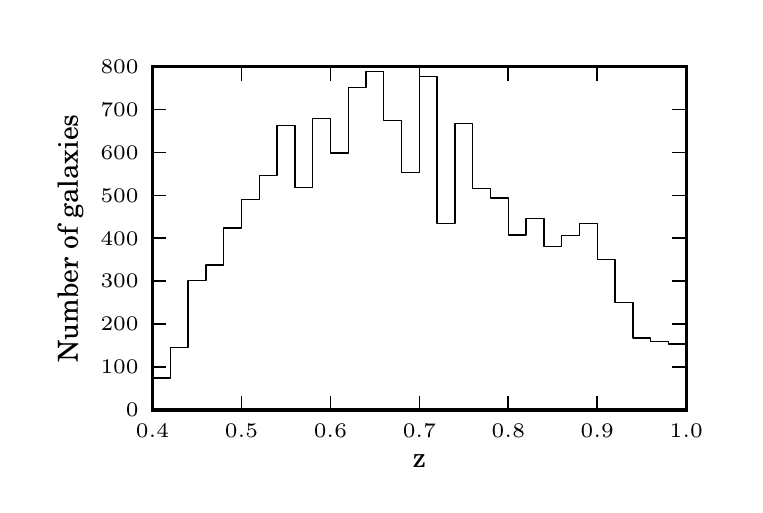}
\caption{\small{The redshift distribution of the 27,350 VIPERS
    galaxies used in this study.  We have limited the redshift range
    of the sample  to $0.4<z<1.0$, and have applied cuts based on
    spectral quality.}\label{fig:zeta}} 
\end{figure}

This study is the first performed on the data of the new VIMOS Public
Extragalactic Redshift Survey (VIPERS), the largest redshift survey
program currently underway at the European Southern Observatory Very
Large Telescope (VLT) (Guzzo et al. 2012, in prep.). VIPERS is designed to map 
in detail large-scale structure over an unprecedented volume of the
$z\sim 1$ Universe.   

In this paper, we develop a specific PCA aimed at analysing
and classifying the spectra collected by the survey. We show that the
technique is capable of compressing the majority of the observed
spectral features into a small number of components, allowing an
objective classification of the vast majority of the spectra in the
sample.  

The reasons for doing this on a survey like VIPERS are manyfold.
First, it represents a way to objectively classify the survey spectra
according to their spectral features.  We shall show in the following
how true this is by analysing both theoretical models and galaxy
templates obtained from observed spectra.  A further, important
motivation is the possibility to homogeneously define sub-populations
of galaxies, to be used for cosmological and evolutionary studies. For
example, the analysis of galaxy sub-samples with different bias
factors provides a way to reduce the impact of cosmic variance on the
measured cosmological parameters (e.g. \citealt{MS2009}). 
A PCA classification can also separate active and passive galaxies, helping to see the effects of environment on galaxy evolution.
Furthermore,
the classification can be used to help identify, in the VIPERS
redshift range, the progenitors of specific populations of galaxies
observed in the local Universe, as the Luminous Red Galaxy sample of
the SDSS (see for example \citealt{Wake2006},  \citealt{Tojper2010},
\citealt{Tojeiro2011}), or for an analysis of correlation functions in
the framework of redshift space distortions \citep{Tojeiro2012}. 

The paper is organized as follows: in $\S$2 we present the data
and reduction steps; in $\S$3 we introduce the Principal
Component Analysis, and the way we implement it as to repair and clean
the VIPERS spectra, along with tests on the
effectiveness of our routines. In $\S$4 we show the
classification obtainable for the VIPERS spectra through this approach
and compare it to the results obtained on stellar population synthesis models.
In $\S$5 we summarize the results.

\section{Data}
VIPERS\footnote{\url{http://vipers.inaf.it}} will target $\sim 10^5$ galaxies for spectroscopy at
redshift $0.5 < z \lesssim 1.2$.  The sample is selected from the 
Canada-France-Hawaii Telescope Legacy Survey Wide (CFHTLS-Wide)
optical photometric catalogues \citep{CFHTLS}. The target sample covers an area
of $\sim24$ deg$^2$ divided over two areas within the W1 and W4 CFHTLS
fields.  Targets are selected to a limit of $i_{AB}<22.5$ and a colour
pre-selection with the $gri$ photometry is used to effectively remove
galaxies at $z<0.5$.  The detailed description of the target selection
can be found in Guzzo et al. (2012) (in preparation). 

The spectra are obtained with VIMOS LR Red grism at moderate resolution ($R=210$).  The
wavelength coverage is 5500-9500$\rm{\AA}$.  The data are processed with
the {\sc Pandora Easylife} reduction pipeline (Garilli \emph{et al.} 2012,
in prep.).  In this work we utilize flux normalized  spectra and
variances as well as masks indicating where spurious  features in the
data have been removed.  

Redshifts and quality flags are measured with the {\sc Pandora EZ}
(Easy Z) package 
\citep{Garilli}.  
The redshift and flag assigned by the {\sc Pandora} pipeline has been
checked and re-fined, for every spectrum, by members of the VIPERS team, ensuring the
reliability of the assignments.

The quality flag indicates the confidence of the redshift measurement
in a similar manner as used in the VVDS \citep{LeFevre05} and zCosmos
catalogues \citep{zcosmos}.  The flag takes the  form $\pm XY.Z$. Negative
values are reserved for spurious, undetected or unidentified serendipitous sources.  The first
digit $X$ indicates the class of object: it is blank for normal galaxies; 1 for
broad-line AGNs, and 2 for untargeted sources serendipitously measured.  The second digit $Y$ indicates the
confidence of the redshift measurement.  Secure redshift measurements
with nearly 95 per cent confidence are assigned $Y=4$.  Measurements with
90 per cent confidence limit are assigned flag 3.  
Flag 2 measurements have been shown to correspond to a confidence limit of
about 80 per cent.  Flag 1 sources are 
highly uncertain at the 50 per cent confidence level, and flag 0 is given
when a redshift could not be assigned.  For this reason, these two classes are not
considered in the present analysis, to guarantee a clean and reliable
sample. Finally, flag 9 is given
to redshift measurements that are based upon only a single emission
line feature. The flag also has a decimal part that indicates the
agreement between the photometric redshift estimate and the
spectroscopic redshift, but we do not use it here.

The total number of VIPERS spectra available for this first study
before any quality cut is 37382, corresponding to the internal data release V2.0 of 23
December 2011.  Our further selection excludes low-quality spectra
as defined above, but includes sources classified explicitly as broad-line AGN 
and secondary sources observed by chance.  We note that there is no harm in including
peculiar spectra as AGN in the overall PCA.  Being rare
cases, these have no effect on the evaluation of the 
Principal Components characterizing the main galaxy sample (see
\S~\ref{sec:pca}).  At the same time, as we shall 
discuss in \S\ref{sec:agn}, it will be interesting to check how
AGN-like spectra can be identified by the PCA as ``outliers'' among
the more standard galaxy spectra.  This may lead also to detection of
more AGN-like spectra,  which do not appear explicitly classified as such. 
\begin{figure}
\includegraphics[scale=1.2]{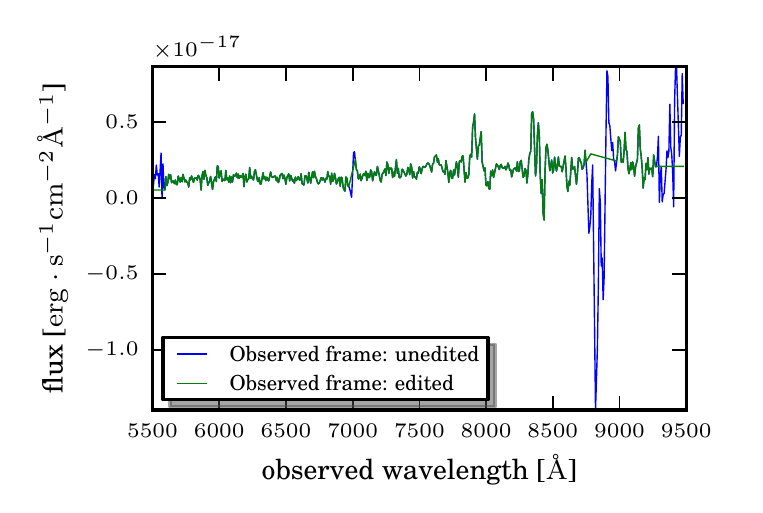}
\includegraphics[scale=1.2]{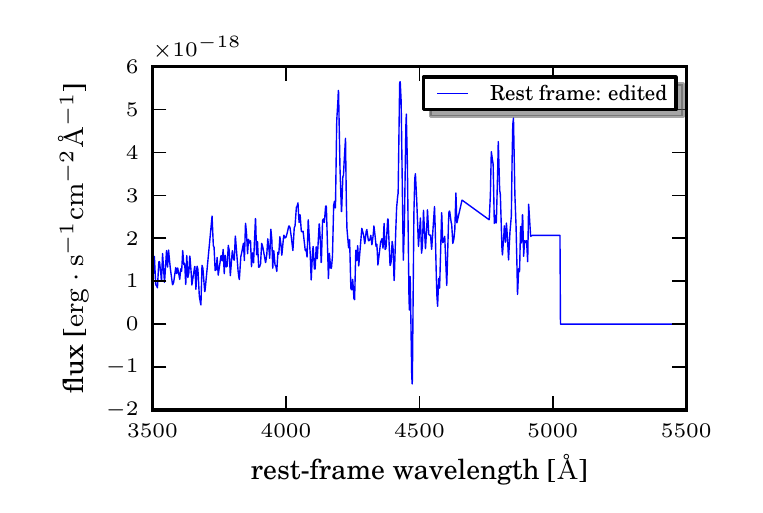}
\caption{\small{\textbf{Top}: huge noise spike (blue line) due to bad sky
    subtraction in a VIPERS spectrum at $z\sim 0.88$, and relative edited spectrum (green line). \textbf{Bottom}: The spectrum
    after resampling on the rest frame wavelength grid.  Spurious
    features have been replaced with a linear interpolation or a flat extrapolation (as in the green spectrum above), and the
    flux has been set to 0 where no data is available at
    $\lambda>5000$, due to the shift to the
    rest-frame.}\label{fig:hole}}
\end{figure}

Since the spectra are observed over a fixed wavelength range, the
spectra must be shifted and mapped to a common rest frame
wavelength scale. We have defined the rest-frame wavelength scale
ranging from $3500<\lambda<5500 \rm{\AA}$, to get the maximum coverage
in all redshift bins. The redshift range is  $0.4<z<1.0$, which covers
a large fraction of the redshift range of the survey,  excluding very
far and very near objects.  
The final sample, after the cuts, includes 27350 spectra
($\sim 73$ per cent of the total in V2.0). 
The resulting redshift distribution of the sources used
in this analysis is shown in Fig. \ref{fig:zeta}. The wavelength
binning we chose to adopt in this work 
increases logarithmically, such that the last interval in the reddest
region has a width of 1$\rm{\AA}$ giving a total number of bins of 2486.
This wavelength scale ensures that every VIPERS spectrum is
oversampled in the rest frame.  The spectra are shifted by a factor of $(1+{\rm
  z})^{-1}$ and resampled with linear interpolation on to the
rest-frame grid.  The variance, given for each spectrum by the square of the relative VIPERS noise spectrum, is
processed in the same fashion.

Necessarily, resampling a spectrum on to the rest-frame grid can leave
gaps at the start or end of the scale, depending on the redshift.
Fig. \ref{fig:hole} (bottom panel) shows a VIPERS spectrum from a high
redshift galaxy after shifting it to the rest frame.  No data is
available at $\lambda > 5000\rm{\AA}$ and in this range the flux is set to 0.
Additionally, the VIPERS spectra are affected by fringing redwards of
8000\AA~induced by the CCD detectors in the VIMOS instrument. The
effect was reduced subsequent to the VIMOS refurbishment in August
2010, and about half of the V2.0 sample used here was
obtained with the old detector. Fringing can leave strong residuals in
the spectrum after sky subtraction, hindering the measurement of
spectral features. In some cases, these spikes have been cleaned in
the reduction/validation phase, and replaced with a linear
interpolation across the spectral region.  The presence of these large
noise artefacts makes the reconstruction of real spectral features
less robust and more complex  (Fig.\ref{fig:hole}-Top).  For our
analysis we would like to develop a procedure to repair these
defects.  We address this through an iterative algorithm that
simultaneously repairs the spectrum and finds the principal
components, as suggested in \citet{CS1999} (see \S 2.1.).  

An important consideration before moving to the analysis is how to
normalize each spectrum. The apparent flux of the source introduces an
arbitrary  
scaling factor that should be normalized out to build a homogeneous
sample. Amongst many possible normalizations, we choose to normalize
each spectrum by a scalar-product normalization, such that for a spectrum $f_\lambda$,
the normalized spectrum becomes $\overline{f}_\lambda=f_\lambda/\sqrt{\sum
  f_\lambda^2}$. The choice is dictated by the fact that normalizing
by scalar product offers advantages for our  classification over other
possible normalizations \citep{Connolly95}: a  normalization based on
morphology would rely on a model distribution of
morphological types in given sample, and may lead to the accidental
suppression of a common galaxy type within the first principal
components of the sample; a normalization by the integrated flux will
give similar results as one done by scalar product, in terms of
principal components, but this second one produces unit vectors representing the spectra, and unit principal components. This means that the coefficients of the decomposition of
each SED on the principal components lie on the surface of an N-dimensional
hypersphere (if we consider N principal components), and thus can be
parametrized by using only N-1 parameters (see \S4.1).   

\section{The Principal Component Analysis}\label{sec:pca}
The Principal Component Analysis (PCA) is a non-parametric way
to extract the majority of information from a noisy dataset, composed of objects which are not completely different one from another. The key characteristic of the PCA in this case is, in fact, the ability to describe a large sample through a reduced number of components, which is guaranteed by the fact that the objects in the sample share many common features (e.g. different measurements of the same quantity, a collection of objects in a catalogue, etc...). This holds true for a sample of galaxy spectra
that are generated by a common underlying physical mechanism, i.e. the radiative physics in the galaxies. 

PCA finds the linear transformation that changes the frame of
reference from the observed or natural one to a frame of reference that
highlights the structure and correlations in the data.  This is done through a rotation of the parameter space such that the axes are aligned along the directions of maximum variance of the data.  This transformation may be found by diagonalizing
the data correlation (or covariance) matrix, whose eigenvectors effectively represent the axes of the new coordinate system.

The basis of the principal components one obtains will be made up by
orthogonal (i.e. uncorrelated) vectors or \emph{eigenvectors} which
are linear combinations of the original variables.  The PCA has the
advantage to describe a set of measurements exploiting dimensions of
the problem which are uncorrelated, and that can be easily ordered by
decreasing importance. This allows us to retain
just a (small) subset of  components, describing
the data using a basis of only a few eigenvectors.

Our goal is to reduce the complexity of a sample of spectra
by expressing them through just a handful of the principal components.
 In particular, we may write an observed spectrum as a data vector containing
$N$ fluxes $f_{\lambda}$, where $\lambda$ indexes the $N$ wavelength bins.
Our sample contains $M$ spectra, and we can write the sample correlation between
wavelength bins as a matrix,
\begin{equation}
C_{\lambda_1,\lambda_2}=\frac{1}{M-1}\sum_{i=1}^M f_{\lambda_1}^{i}f_{\lambda _2}^{i}\label{eq:corr},
\end{equation}
where $i$ indexes the spectra in the sample and $\lambda_1$ and
$\lambda_2$ index wavelength bins.
The correlation matrix can be decomposed into a set of orthonormal eigenvectors, or \emph{eigenspectra} $e^i_{\lambda}$
and eigenvalues $\Lambda_i$,
\begin{equation}
C_{\lambda_1,\lambda_2}=\sum_{i=1}^M e^i_{\lambda_1}\Lambda_i e^i_{\lambda_2}.
\end{equation}
The eigenspectra are ordered with decreasing eigenvalue such that the most common
features within the spectra are contained in the first few eigenspectra.

The eigenspectra form an orthogonal basis or \emph{eigensystem} and any spectral energy distribution, $f_{\lambda}$, can be expressed as a
 sum of the $M$ eigenspectra with linear coefficients $a_i$:
\begin{equation}
f_{\lambda}=\sum_{i=1}^M a_i e^i_{\lambda}.
\end{equation}
Since the higher eigenspectra carry little statistical information about the spectra
we may truncate the sum to use only the first $K\ll M$ components.  We refer to this
as the reconstructed spectrum  $\hat{f}_{\lambda}$,
\begin{equation}
\hat{f}_{\lambda}=\sum_{i=1}^K a_i e^i_{\lambda}\label{eq:pca},
\end{equation}

The correlation matrix, as defined in (\ref{eq:corr}), will have dimension given by the number of
wavelength bins (2486x2486).  In the literature, it is also common to
define the correlation matrix such that the dimension is the number of
spectra \citep{Connolly95}.  This is clearly inefficient when the number of
spectra is greater than the number of wavelength bins.

An additional result obtainable by the PCA projection of eq. (\ref{eq:pca}) is a measure of the signal-to-noise ratio for each spectrum, as 

\begin{equation}
\frac{S}{N}(f_{\lambda})=\sqrt{\sum_{\lambda}\Big(\frac{\hat{f}_{\lambda}}{\overline{n}_{\lambda}}\Big)^2}\label{eq: sn}
\end{equation}

where $\overline{n}_{\lambda}$ is the VIPERS normalized noise spectrum,
relative to the spectrum $f_{\lambda}$. Given the VIPERS noise spectrum $n_{\lambda}$, the normalized noise spectrum is given by $\overline{n}_{\lambda}=n_{\lambda}/\sqrt{\sum f_{\lambda}^2}$.

\subsection{Repairing bad spectral regions}\label{sec:rep}
\begin{figure}
\includegraphics[scale=1.2]{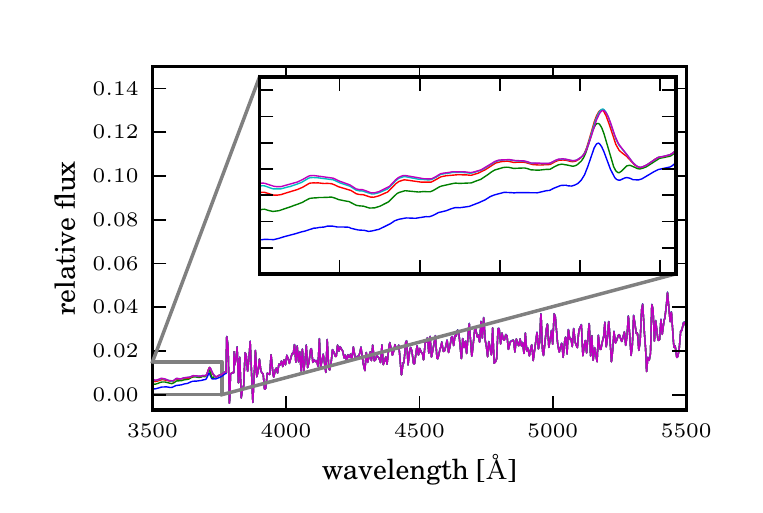}
\caption{\small{A VIPERS spectrum presenting a gap on the blue side, due to rest-frame shifting. The missing data is reconstructed through an iterative routine.  The first five steps (zoomed in the box) go from the first (bottom line) to the fifth iteration (top line).
}}\label{fig:steps}
\end{figure}
A spectrum can be corrupted by instrumental artefacts. VIMOS has its
own specific features, such as the zero-order image from a bright object
in the slit above, or residuals remaining after the subtraction of sky
lines.  In some cases, artefacts have been removed from the spectra by
the reduction pipepline or manually, and have been replaced by linear
interpolations, creating ``gaps" in the spectra, i.e. regions where flux data was lost. Fig. \ref{fig:hole}
illustrates a spectrum with a bad region that has been removed. This
must be properly taken into account when applying a PCA decomposition,
to avoid treating some bad features or noise artefacts as physical
peculiarities that will influence the shape of the eigenspectra, and
hence the whole analysis. 
To do that, we assign a weight to each spectral bin
\begin{equation}
w_{f_{\lambda}}=\frac{1}{\overline{n}_{\lambda}^2}.
\end{equation}
The weight is set to 0 within the gaps and in regions of the spectra
that have been manually edited.  The weight mask is essential to
derive accurate eigen-spectra from data containing gaps. In fact, with a
naive application of PCA to these ``gappy" spectra, it is no longer possible to
construct a set of orthogonal eigenspectra \citep{CS1999}.  We have therefore
developed an algorithm to simultaneously repair the gaps in the
spectrum and compute orthogonal eigenspectra.

At the start of the repairing routine, the gaps in the spectra are
replaced by linear interpolations.  Although, for gaps at the start or
end of the spectrum, we find that it is sufficient to simply set the
flux to 0. We then proceed in an iterative manner.  First, the
correlation matrix is constructed from the spectra and the
eigenspectra are computed.  We keep only the 3 most significant
eigenspectra to perform the following repairing steps.  The
choice of the number of eigenspectra, as discussed later in \S 3.3, is
dictated by the need to be able to describe all the spectra in the
sample, while avoiding the noise, which is reflected by the
eigenspectra from the fourth on.  

We compute the set of eigencoefficients, $\{a_i\}$, for each
spectrum, $f_\lambda$, with a least squares minimization routine.  The
objective function to be minimized is given by,
\begin{equation}
\chi^2=\sum_{\lambda} w_{\lambda}(f^{(i)}_{\lambda}-\sum_{j}a_j e_{j\lambda})^2.
\end{equation}
where $f^{(i)}$ is the spectrum data vector on the
$i$\textsuperscript{th} iteration, $e_{j\lambda}$ is the set of
eigenspectra and $w_{\lambda}$ is the weight vector.  The minimization
is carried out with the Levenberg-Marquardt algorithm implemented in
the Python Scientifical Library 
({\sc Scipy})\footnote{\url{www.scipy.org}}.

We found that in some cases the best-fitting coefficients did not
represent physical spectra.  For example, the continuum of the
repaired spectra could go negative or strong emission lines could be
inverted.  These poor results are usually found for very noisy spectra
or for spectra that are more than 50 per cent masked: in fact, when many 
spectra have been masked in the same range of  
wavelengths, the PCA process is unable to find the 
information to repair the gaps. 
In our VIPERS sample, there are 57 spectra that are missing more than 50 per cent of the wavelegth coverage, while the average gap fraction for the sample is $\sim 10$ per cent.

The other possibility is that some
peculiar piece of 
information needed to recover a spectrum is not reflected within the
chosen eigenspectra (see $\S$2.3).  

These problems in the majority of the cases cause the PCA to fail to reproduce simultaneously the continuum and the line features of these spectra, leading usually to the inversion of some lines: the continuum pixels have more weight than those in the lines, and the PCA routine reproduces them as accurately as possible at the expense of the line features.
To avoid these degenerate solutions, we introduced a check within the wavelength range of the line features that mostly suffer from this problem in our routine ([OII], H$\beta$ and [OIII]). Whenever the least-square repairing routine finds an inverted line as a solution for the fitting problem, we add an exponential penalty term to the $\chi^2$:
\begin{equation}
\chi^2=\chi^2+c*\sum_{l}e^{(D_{l}-D_0)/D_0}
\end{equation}
where $c=$2486 is the number of bins in a spectrum, $D_{l}$ is the difference
between the continuum and the line peak for each line $l$, and $D_0=$0.005 is the threshold above
which the penalty is applied. The value of $D_0$ has been chosen such
a way to impede the PCA to reverse emission lines, whilst avoiding
this penalty to be applied by small real throats within the elected
wavelengths, for example in red galaxy spectra.  In this way, whenever
the PCA finds a negative solution for a real emission line during the
phase of repairing, the $\chi^2$ gets raised and the routine is
therefore forced to find a set of eigencoefficients corresponding to a
more physically realistic reparation. The specific choice of this
shape for the penalty has been the result of a number of  tests using
different functions, given the freedom allowed by  the problem.

After finding the best-fitting coefficients, $\{\hat{a_i}\}$, we
reconstruct the spectrum as,
\begin{equation}
y_\lambda = \sum_{j}\hat{a}_j e_{j\lambda}.
\end{equation}

\begin{figure}
\center
\includegraphics[scale=0.3]{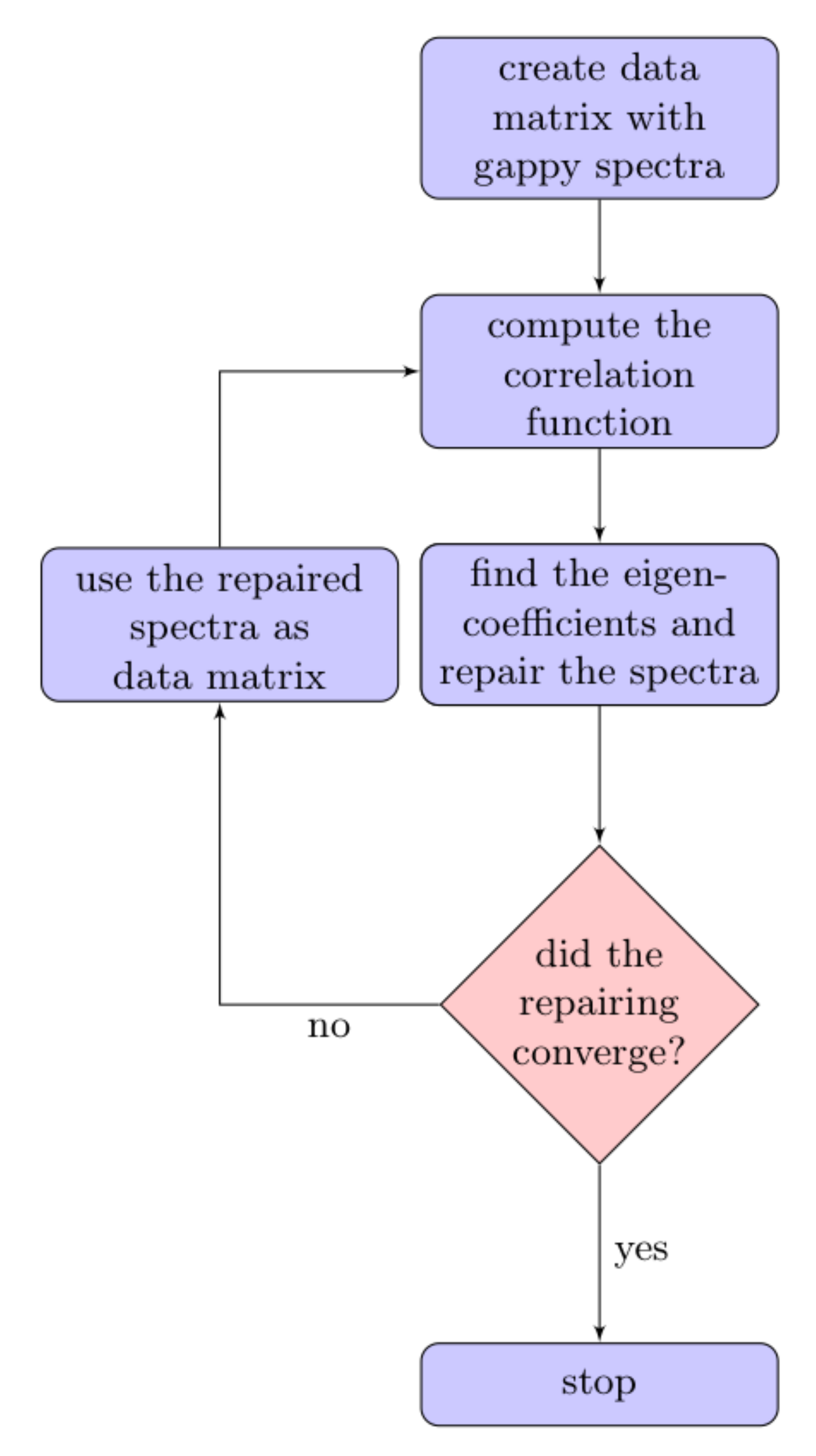}
\caption{\small{Flow chart of the PCA repairing process.}\label{fig: flowchart}}
\end{figure}

We then replace the gaps (and only the gaps) in the original spectrum with portions of the
projection.  In Fig.\ref{fig:steps} we show an example of
different stages of repairing.  At each iteration the spectra are
renormalized by their scalar products (the normalization changes
because the gaps are updated on every loop). The routine
progresses as shown in the diagram of Fig (\ref{fig: flowchart}).
Once the eigenspectra of the repaired galaxy sample are obtained
(Fig. \ref{fig:esp}), we can project each spectrum on to the eigenbasis
to get the set of eigencoefficients $a_i$.

The convergence of the routine is safely reached, for each of the spectra, within the twentieth iteration of the process, when any further refinement of the value of the eigencoefficients for the repairing does not change the repairing significantly, as shown in \S3.2.

\subsection{Tests with mock spectra}\label{sec:tests}
To test our routine we created a synthetic sample of galaxy spectra.
The spectra were generated using two sets of templates: a subset of
the Bruzual\&Charlot (B-C hereafter) \citep{BC2003} model spectra
(which do not contain emission lines), to obtain realistic
early-type galaxies, and the 12 Kinney-Calzetti templates (K-C hereafter)
\citep{Kinney96, Calzetti94}, covering from pure bulges to starburst
galaxies, to give a total of 45 template spectra. We computed the first five eigenspectra of these templates
to define an orthogonal basis spanning the range of galaxy types. We then
constructed mock spectra that are similar to the templates by
generating Gaussian distributed numbers as eigencoefficients. This Gaussian distribution is centered on the first 5 eigencoefficients of the starting template set, with variance given by the relative eigenvalues.
We generated 450 mock spectra around
each template giving a total sample of 20,250 spectra, that reduces to about 16,000 once spectra presenting unphysical features (i.e. inverted emission lines) are removed. 

\begin{figure}
\includegraphics[scale=1.2]{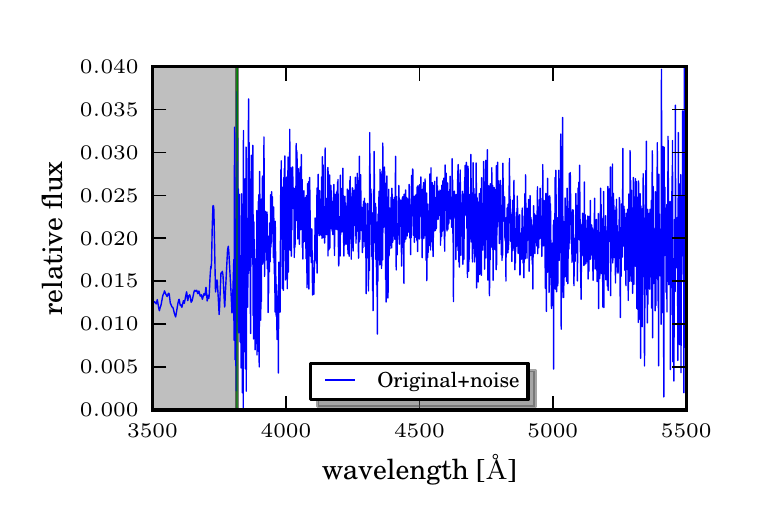}
\includegraphics[scale=1.2]{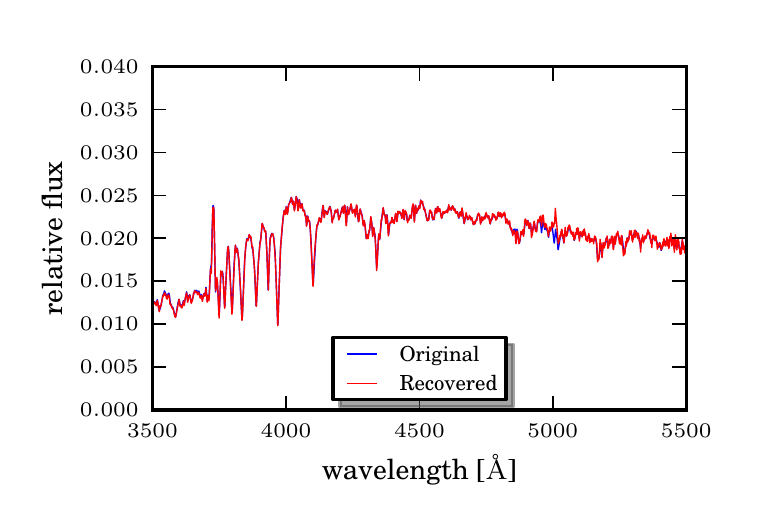}
\includegraphics[scale=1.2]{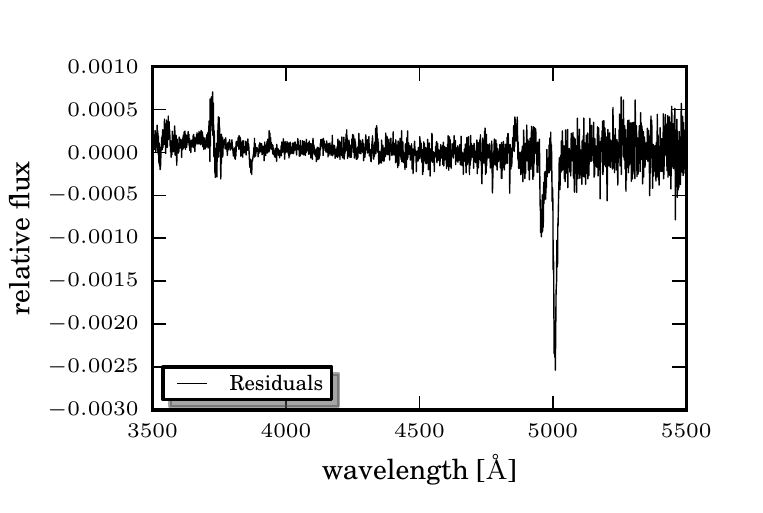}
\caption{\small{\textbf{Top:} a synthetic spectrum with synthetic noise added.  The shaded region would be masked and reconstructed. \textbf{Middle}: qualitative comparison between the original spectrum before the noise has been added (blue) and its reconstruction through the PCA routine (red). \textbf{Bottom}: residuals between the mock and its reconstruction. The possible differences between the intensities of the real and the recovered emission lines are acceptable for our classification system, since it is more sensitive to the continua of the spectra than to the line features.}\label{fig:synth}}
\end{figure}

We next degrade the spectra with synthetic noise to simulate the
VIPERS data.  Each synthetic spectrum is assigned the same data variance
and weight mask of a randomly selected VIPERS galaxy. The synthetic
noise spectra are generated from a Gaussian realisation with the associated VIPERS variance, as illustrated in the top panel of Fig.\ref{fig:synth}, and
the mask is applied to reproduce the gaps.  In this way, we produce an
artificial data set that can be used to test the fidelity of the
reconstruction procedure.

\begin{figure}
\includegraphics[scale=1.2]{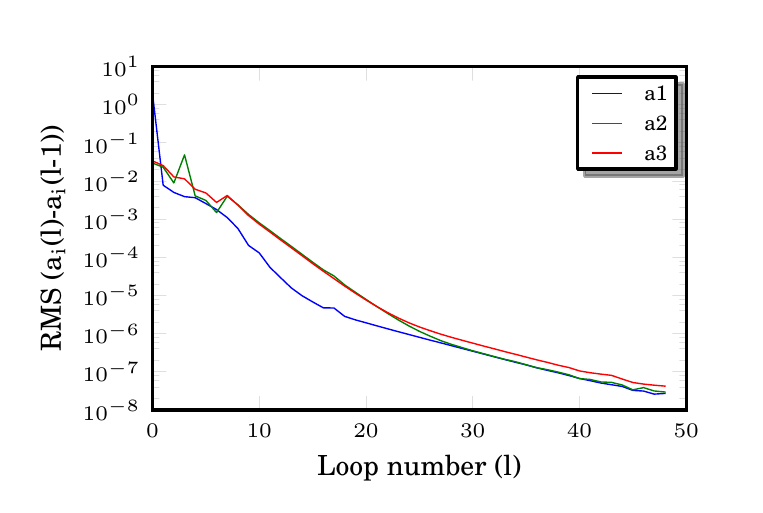}
\caption{\small{The root mean square difference between the eigencoefficients and themselves at the previous iteration, for the repairing of the synthetic spectra. 
The RMS difference steadily decreases on subsequent iterations.
}
\label{fig:tr}}
\end{figure}

We apply the PCA repairing routine with three eigenspectra.  Then we
project the spectra on them, to clean from noise and be able to
compare the recovered spectra to the noise-free synthetic ones. Apart
from slight differences in the intensity of the emission lines (as anticipated in \S\ref{sec:rep}) the
reconstruction is qualitatively good, even where the region to be repaired was a line feature (Fig.\ref{fig:synth}: middle-bottom,
Fig. \ref{fig:tr} for a more quantitative check) . The fit can be improved
by adding more components to the PCA, but, as will be discussed later,
the 4th eigenspectrum is already affected by noise for the VIPERS
sample, and the reconstruction obtained with three is sufficient for
the classification system.

The PCA routine has been run on the synthetic spectra for a large
number of iterations, that we chose to be 50. By looking at the root
mean square difference between the eigencoefficients at each iteration
(Fig. \ref{fig:tr}) we see that the routine is converging: in
particular, the differences between the eigencoefficients become
steadily smaller. The effects of this on the repairing is actually
negligible after five iterations, so we halt the code when
the difference between the eigencoefficients at consecutive loops is
$\le 10^{-3}$, since under this threshold further refinement has a negligable
effect on the results. We found that the repairing for every single spectrum has surely reached $10^{-3}$ within the 15th iteration for $a_1$, at the 17th
for $a_2$ and at the $16$th for a$_3$, so in this case 17 iterations
are enough to repair and recover the original spectra for the
synthetic spectra. To be on the safe side, we decide to take 20 iterations.

\subsection{VIPERS spectra: repairing and cleaning}
We now apply the PCA routine to the VIPERS sample.  As anticipated in
sections $\S\ref{sec:rep}$ and $\S\ref{sec:tests}$, we must decide on a stopping point for
the repairing routine and the number of eigenspectra to use.

\begin{figure}
\includegraphics[scale=1.2]{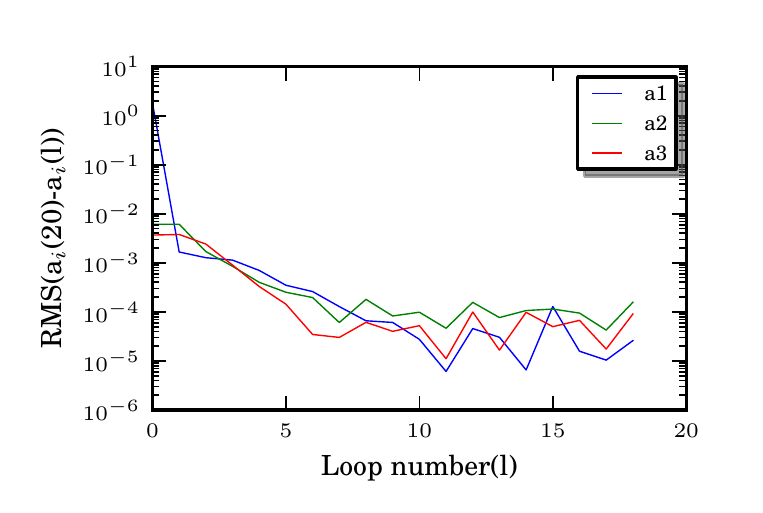}
\caption{\small{The RMS error on coefficients for VIPERS spectra.  Plotted is the root mean square difference of the coefficients of the decomposition after 20 iterations, and themselves at the $i^{\rm th}$ iteration. For a particular spectrum the difference actually starts oscillating around 0 with decreasing amplitude after the 5-10th iteration on average.}\label{fig:diffconv}}
\end{figure}

As suggested by the tests on mock spectra, we halt the repairing
procedure after 20 iterations.
We may estimate the relative error in the coefficients after each
iteration by measuring the root mean square difference between the
value at iteration $i$ and iteration 20.  Fig. \ref{fig:diffconv}
shows that this error is oscillating at the level of $10^{-4}$ by the 10${\rm ^{th}}$ iteration.

\begin{figure*}
\includegraphics[scale=1.5]{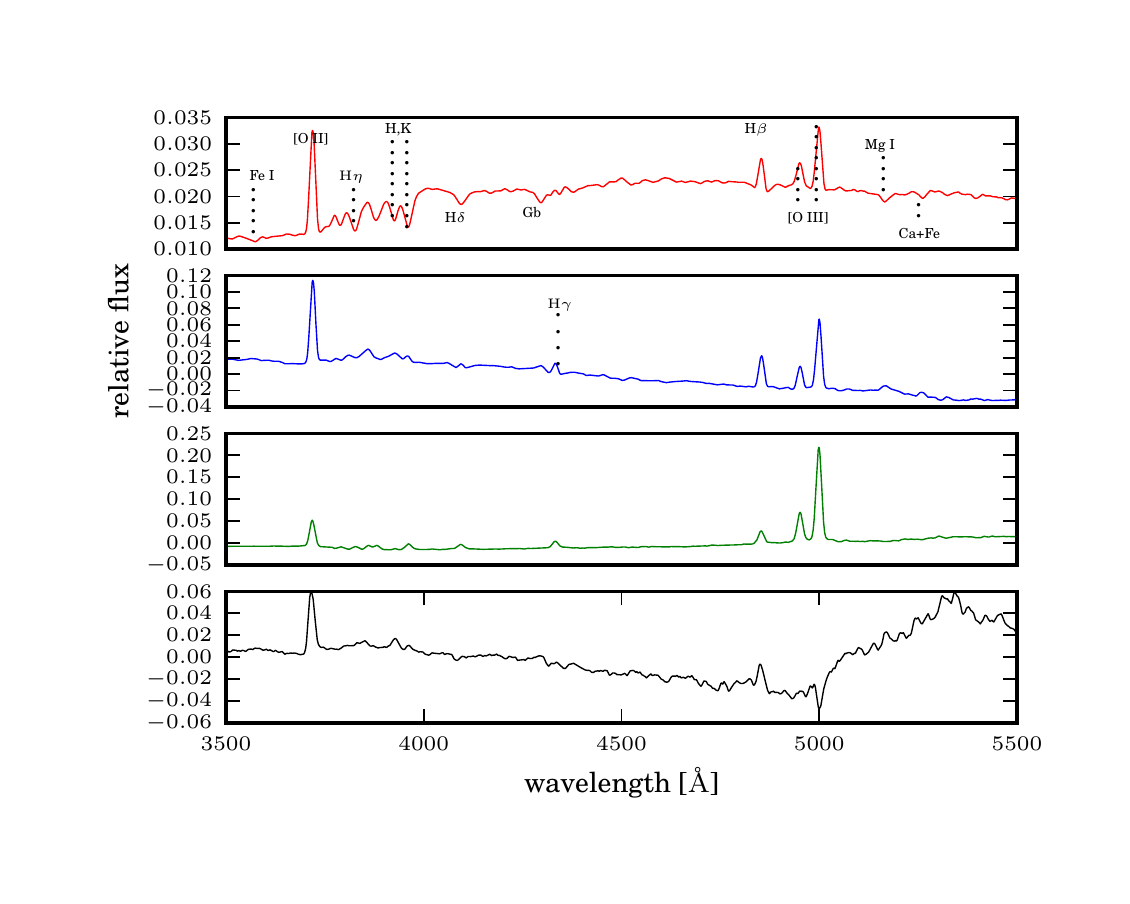}
\caption{\small{The first four VIPERS eigenspectra computed after repairing. From top to bottom the power is decreasing (the first eigenspectrum is at the top, the fourth at the bottom).  The first eigenspectrum mirrors the average of all the spectra, while the second and the third are residuals form the average. Some of the most common spectral features present in the eigenspectra are highlighted in the first eigenspectrum. Systematic effects in the spectra begin to be visible in the fourth spectrum at $\lambda>5000$\AA}\label{fig:esp}}
\end{figure*}

We use three eigenspectra in the repairing procedure to reconstruct
the spectra inside the gaps.  This number should be chosen to be large enough
such that the repairing can reproduce the signal without adding
spurious noise, although the results are not strongly dependent on the
exact number used.

After the convergence of the repairing process, we obtain the complete eigenspectra for the VIPERS
sample.  The first four eigenspectra ordered by significance are shown
in Fig. \ref{fig:esp}.  The first three VIPERS eigenspectra, as shown,
contain the large majority of information on the sample, particularly
the first one, which mirrors the average of all the spectra, while the
others represent the residuals from the mean. In particular, the shape of the
continuum of the first eigenspectrum is comparable to the one of an early-type
galaxy, while it contains also emission lines typical of a star-forming galaxy. The second one instead can be associated to a late-type spectrum,
while the third can be thought of as an intermediate galaxy SED. The
fourth one, at $\lambda<4500$\AA, adds information about the
intensity of the [OII] emission line and the continuum resembles the one
of a blue galaxy, but redward of 4700 \AA~ it shows an unphysical bump
that is not expected in a galaxy continuum.  We attribute this to the fact that, redwards of  $\lambda_{obs}>8000$\AA~, VIPERS spectra are affected by systematic effects arising from the coupled effect of detector fringing and strong sky emission lines (Guzzo \emph{et al.} 2012, in prep.). For low signal-to-noise objects the repairing of this region is probably more affected by systematic uncertainties that can heavily influence the PCA reconstruction. 
Thus, to
effectively repair the spectra without spurious features, we use
only the first three eigenspectra.

By a simple estimate of the power enclosed in each eigenspectrum
\begin{equation}
P(e_i)=\frac{\Lambda_i}{\sum_{i=1}^{tot}\Lambda_i},
\end{equation}
where $\Lambda_i$ are the eigenvalues of the correlation matrix,
we find that the first three eigenspectra hold $\sim$ 90.6 per cent of the
total power; the first contains $\sim$ 87.3 per cent, the second $\sim$
2.5 per cent, the third $\sim$0.7 per cent, and from the fourth on the power
content starts to decrease rapidly with respect to the first three, see Table \ref{power}.
The variance in each component is a measure of the information
content and we can conclude that three eigenspectra are
enough to describe the sample in a statistical sense.  However, we
will see that this measure of information does not translate directly
to the physical information contained in spectral features, as anticipated in \S\ref{sec:rep}.  For
example, we found that the slope of the continuum is well described
by just a few eigenspectra, but this is not true for the line
features.  The information on the lines in some cases is contained into higher-order
components, that we neglect to avoid the noise, even though we recognize that this information
is essential for understanding the physical properties of galaxies.

\begin{table}
\begin{tabular}{|l|r|}
\hline
Power of the first three eigenspectra&$\sim 90.56\%$ \\
\hline
First eigenspectrum&$\sim 87.30\%$ \\
Second eigenspectrum &$\sim 2.54\%$ \\
Third eigenspectrum&$\sim 0.71\%$ \\
\hline
Fourth eigenspectrum&$\sim 0.17\%$\\
\hline
\end{tabular}
\caption{The power contained in the first four eigenspectra.\label{power}}
\end{table}

\begin{figure}
\includegraphics[scale=1.2]{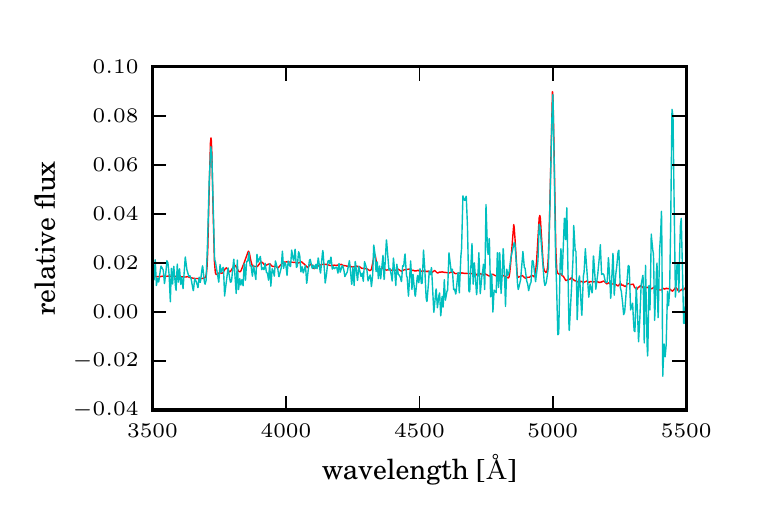}
\includegraphics[scale=1.2]{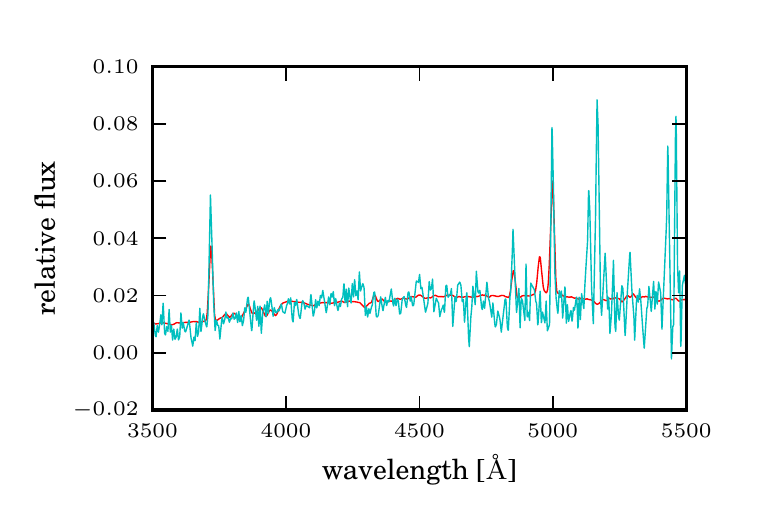}
\caption{\small{Two repaired and cleaned VIPERS spectra (red) superposed to themselves after the only repairing process (cyan). Our projection method is statistically able to recover the realistic emission and absorption features together with the slope of the continuum. This is a consequence of the combination of ``cleaning", operated by the description of the spectra through the first three eigenspectra, which do not reflect the noise of the sample, and least-square fitting with introduction of penalty terms in the regions of the lines.}\label{fig:proj}}
\end{figure}

After the repairing process, by projecting the VIPERS spectra on to the basis of three final eigenspectra
we can achieve our goal of cleaning the spectra from noise, as
illustrated in Fig. \ref{fig:proj}. This is guaranteed by the fact that the first three eigenspectra are affected very little by noise. The same simplification offered by the PCA in using only three components makes it impossible, though, in our specific case, to naively apply Eq. (\ref{eq:pca}) to recover properly VIPERS spectra. In fact, as for the repairing process, the projection on to only a few components is
not guaranteed to reproduce spectral features matching the data. And again, as for the repairing,  the projection can invert lines or add lines not present in
the data.  These errors arise because additional components are needed
to recover all the lines accurately. We find that 5 per cent of spectra show
unphysical line feaures once projected on to 3 components only. The
situation can be improved by adding more components to the projection; however, this will re-introduce noise and artefacts,
again degrading spectral features.

We can arrive at a compromise by assigning greater importance to the
physical recovering of emission lines.  This is precisely what was
done in \S\ref{sec:rep} where penalty terms were added in the least-squares
minimization procedure to find the best-fitting, but physical
repairing.  We adopt this routine again in the final step to
project each spectrum.  The safeguard of the physicality of spectra is constrained imposing
that the continuum is positive and the [OII], H$\beta$ and [OIII] lines are not
inverted. 
By comparison of the equivalent width of the [OII] line in the repaired and projected spectra to the same feature in the original spectrum, we find that the line, on average, is recovered with a precision of $\sim 20$ per cent, whereas for $\sim 68$ per cent of the spectra the line is recovered within 10 per cent. This is in agreement with the results found by \citet{Yip2004} for the majority of SDSS spectra in their analysis with 3 eigenspectra. For the reconstruction of the problematic emission line spectra only, they chose instead to use 10 eigenspectra, obtaining an error on the recovering of the lines of order 15-25 per cent. 
Finally, the final quality of the repairing in our analysis, after the penalty has been applied, doesn't show any clear correlation to the portion of gaps in a spectrum, even if larger gaps easily increase the possibility of unphysical reconstructions at first step.

\section{Classification of VIPERS spectra}

\subsection{K-L Projection}

\begin{figure*}
\includegraphics[scale=1.5]{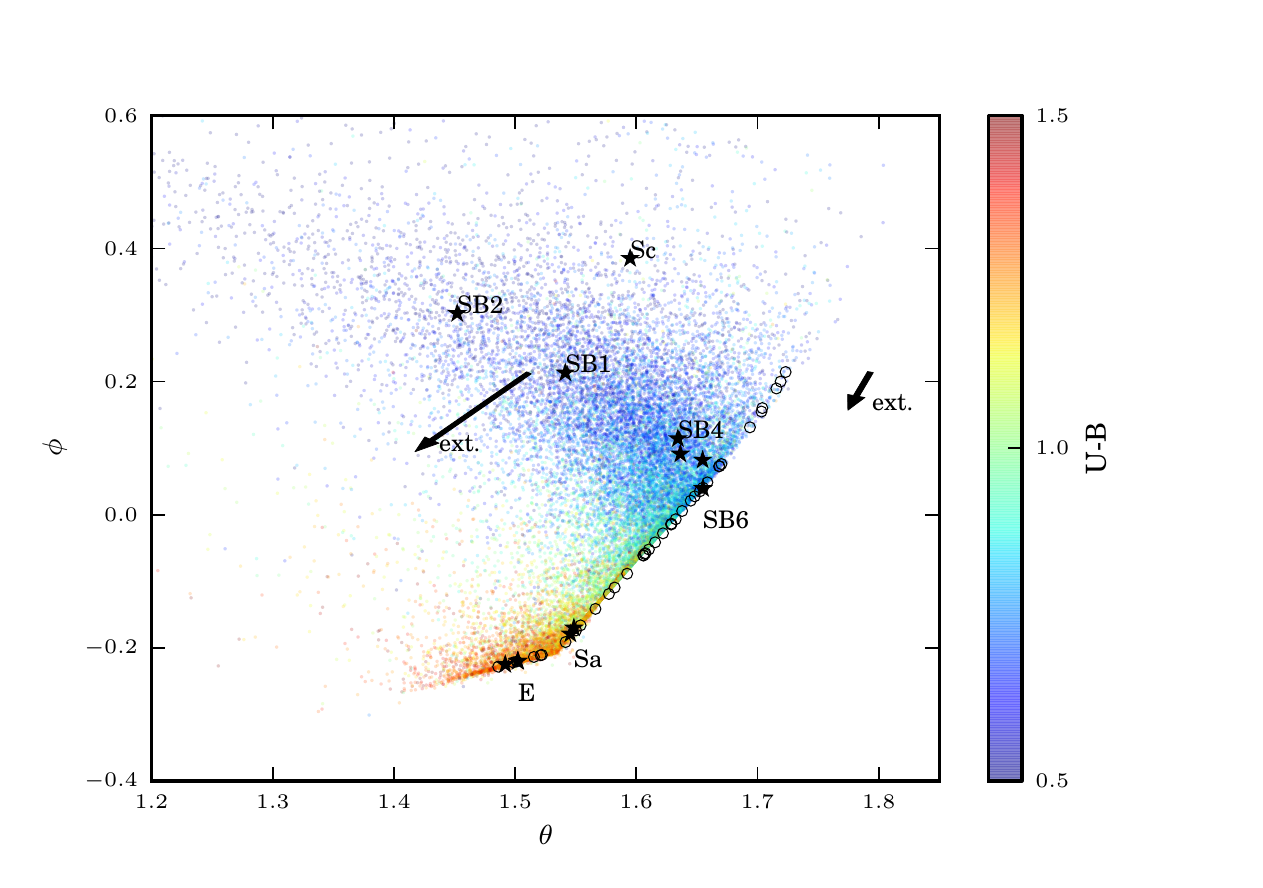}
\caption{\small{The K-L plot, $\phi$ versus $\theta$, for VIPERS repaired and cleaned galaxies, with the position of Bruzual-Charlot and Kinney-Calzetti model galaxies overplotted. The colour gradient of the points from red to blue through green represents the $U-B$ rest frame color of each galaxy in the sample.
The sequence
of circle markers represents the B-C models ranging
from the reddest (early-type) to the bluest (late-type) continuum slopes.  The Kinney-Calzetti templates (star markers) are labelled with galaxy type.  The early type galaxies are positioned with the early-type B-C templates, while the  starburst templates are found in the middle.  The sharp edges in the distribution on the right hand side arise from constraints applied in the PCA reconstruction.  Finally, the arrows show the effects of dust extinction for the two sets of models, with $A(V)$=1 mag and $R_{V}$=3.52.}\label{fig:pt}}
\end{figure*}

The eigen-coefficients $a_1$, $a_2$ and $a_3$ form an optimal basis in
which to classify the spectra. To further reduce the parameter space to a non-degenerate basis we compute the Karhunen-Lo\`eve angles (K-L hearafter) \citep{Connolly95, Karhunen47, Loeve48}, so defined:
\begin{equation}
\phi=\tan^{-1}\Big(\frac{a_2}{a_1}\Big)
\end{equation} 
\begin{equation}
\theta=\cos^{-1}a_3
\end{equation}
The two angles $\phi$ and $\theta$ fully parametrize the three dimensional space
because, owing to the normalisation constraint, the coefficients fall
on the surface of a 3D sphere. 

\begin{figure}
\includegraphics[scale=1.2]{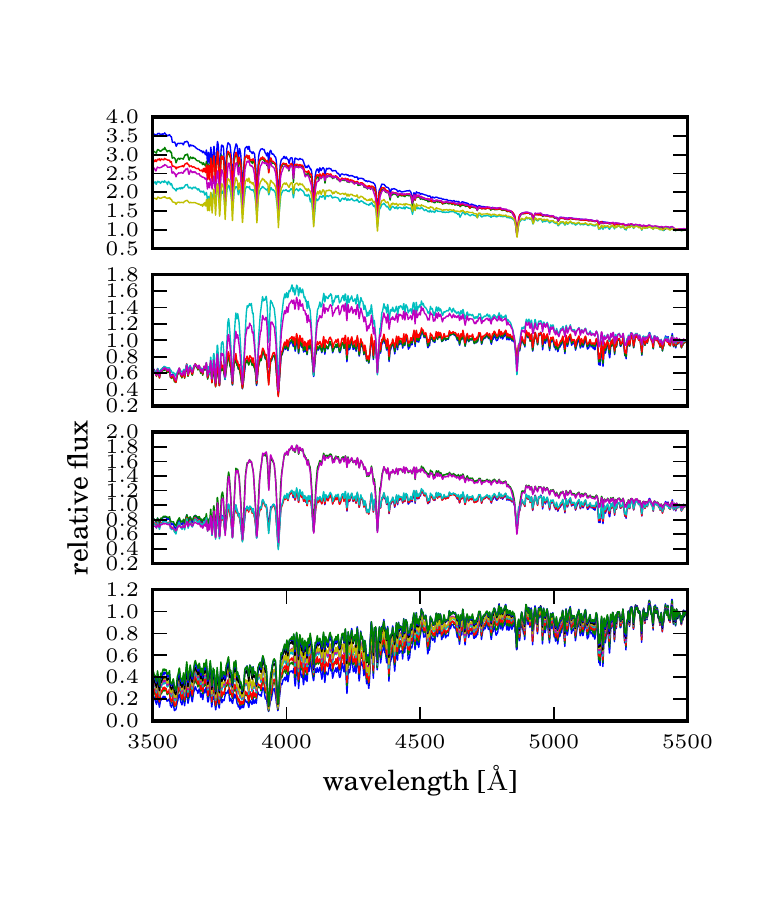}
\caption{\small{The B-C spectra corresponding to the circles in
    Fig.~\ref{fig:pt}:
    red templates (bottom) lie in the low-$\phi$ region, with 
    intermediate templates instead occupying the range
    $-0.2<\phi<0$ (middle boxes), and bluer ones lying at the top of
    the K-L plot.}\label{fig:move}} 
\end{figure}

To pin down the location of different galaxy types on the
$\phi-\theta$ plane, we take advantage of the same group of B-C model spectra from which we picked the templates used to
test the repairing routine (keeping also the blue galaxy
representatives, although these are not fully realistic because of the
lack of emission lines). We project them on the three VIPERS
eigenspectra and then obtain the K-L angles, which are shown in Fig. \ref{fig:pt}.

We find that in the K-L plot, the redder galaxies lie towards negative
values of $\phi$ and quite small values of $\theta$, while, as $\phi$
and $\theta$ increase, the galaxies become bluer (Fig.
\ref{fig:move}), as suggested by the $U-B$ rest-frame color of VIPERS galaxies. Since an increase in $\phi$ is equivalent to an
increase in $a_2$, this means that the bluer galaxies are represented
by larger values of $a_2$ (and viceversa for the redder ones). This
is expected, since the shape of the second eigenspectrum is the one
that most resembles the spectrum of a blue galaxy. We do not consider
now the first eigencoefficient $a_1$, because, being related to the
first eigenspectrum, which is the average of all the spectra, it is
not a significant discriminator. 
Let us remark again, though, that we are basing this interpretation on a set
of model spectra that do not present emission
lines, although they do trace the continuum of blue galaxies in some cases. So
they give a general idea of the arrangement of different spectral
types on the K-L plot, but they are not apparently able to span the
full distribution.

To get more quantitative information on how galaxies spread on the K-L
plane, we performed the same comparison using the Kinney-Calzetti
templates (Fig. \ref{fig:pt}). These are the same we used in \S2.2 to build the synthetic
spectra for the test, together with the B-C red-intermediate
spectra.

The K-C templates provide confirmation that the earliest
type galaxies are at the bottom of the K-L plot, as suggested by the
bulge and elliptical K-C templates.  Additionally, the K-C-Sa and K-C-Sb spiral
galaxies fall near to the region of intermediate B-C models, consistent with them presenting a certain level of star formation. The
starburst galaxies, instead, follow a branch which is nearly
orthogonal to the trend followed by red and intermediate
galaxies. Finally, the K-C-Sc template occupies the highest position
in $\phi$ in the plot, due to the steepness of its continuum, and it is more shifted towards lower values of $\theta$ with respects to B-C models, due to the presence of emission lines. 
We also found that moving towards lower values of
$\theta$ corresponds to increase the intensity of emission lines; this will become evident in section \S\ref{sec:groupfinding}. So we can state that the two K-L parameters $\phi$ and $\theta$ are related to the age and to the star-formation-rate in a rather complex way: an age sequence can be observed moving along the direction of the ridge of normal galaxies, at the right edge of the K-L plot, while an instantaneous star formation sequence can be observed on the perpendicular direction.

The sharp bottom and right edges of the cloud of data points in
the K-L plane are a consequence of the least-square penalty terms,
introduced in the projection of the sample over the eigenspectra
basis, together with the limits imposed by our two-components parametrization.  These two boundaries limit forbidden regions beyond which the
reconstructions would be unphysical with negative continua or inverted
emission lines due to the possible lack of information of the chosen components, if the penalty was not applied. Consequently, spectra with no emission lines are
found at these edges of the cloud of points, as demonstrated by the B-C models.

\subsubsection{Comparison to SDSS data}
\begin{figure}
\includegraphics[scale=1.2]{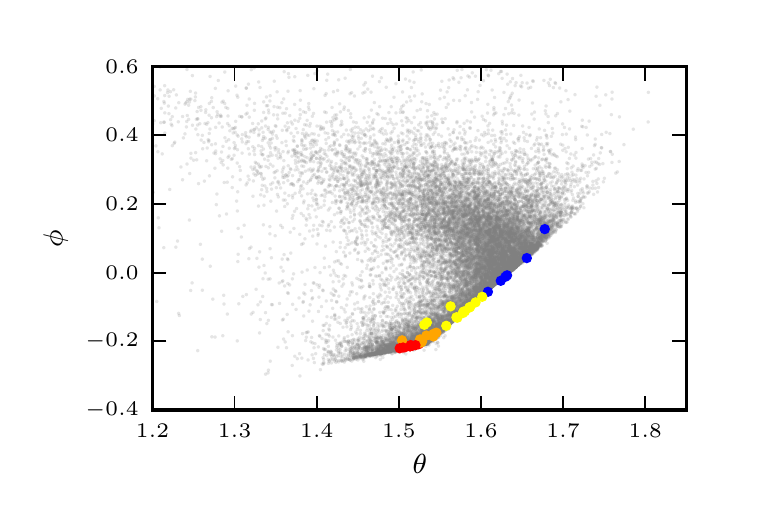}
\caption{\small{The set of 38 SDSS templates by \citet{Dobos2012} as projected on the VIPERS eigenspectra. The templates roughly follow the evolutionary track marked by the right edge of the K-L plot, apart from 3 templates that present stronger emission lines in the red part.}\label{fig:sloan}}
\end{figure}
We can compare the distribution of VIPERS galaxies to SDSS galaxies on
the K-L plot. To this purpose, we used a set of 38 SDSS templates
computed through a PCA projection by \citet{Dobos2012}. The templates
were first re-binned on the same wavelength scale of VIPERS data,
and normalized through their scalar product. They were then simply
projected on the VIPERS first 3 eigenspectra with the same routine
discussed earlier. 

The SDSS templates fall in the region at the right
edge of the plot, following the same track found for the other
datasets. In particular, the majority of them can be found near to the right sharp edge, because their PCA projection over the VIPERS first 3 eigenspectra was finding unphysical solutions for the line features and needed the $\chi^2$ penalty to be applied. The colour gradient, from red to blue, gives a qualitative idea of the colour of the relative template (Fig. \ref{fig:sloan}).
Only a group of three spectra seem to detach from the main branch,
positioning in a region of slightly smaller $\theta$. The reason for
that, as expected, is that those spectra present slightly stronger
emission lines, mainly in the red part, than all the other SDSS
templates. Again, the PCA proves much more sensitive to the slope of
the spectra than to emission lines in positioning the objects on the
$\phi$ scale.  In fact, although the blue templates present strong
emission lines, their slope is flatter than many VIPERS blue galaxies,
causing the templates to hardly reach large numbers in $\phi$.

\subsection{Effect of dust}

A natural question we can now ask about our classification regards the
effects of dust extinction on the position in the K-L plot. To this end, we applied an extinction law to the model templates. Since our purpose is only to check the direction to which extinction moves the galaxies in the K-L plot, we
chose to apply the same simple Cardelli-Clayton-Mathis extinction laws \citep{CCM} to all galaxy types, over the optical-near infrared wavelength
range ($3000\mathrm{\AA}\le\lambda\le 9000\mathrm{\AA}$), which contains the rest frame range we considered for our VIPERS data. The
parameter $R_V[=A(V)/E(B-V)]$, with $A(V)$=1 mag, was set to 3.52.  The
extinction effects on the B-C and Kinney-Calzetti models are
represented by the arrows shown in Fig. \ref{fig:pt}.

Once the B-C models have been corrected for dust-extinction, they all shift towards the bottom of the K-L plot (Fig. \ref{fig:pt}), in the same direction marked by the B-C curve. 
This is consistent with a reddening of the continuum.
 For the Kinney-Calzetti templates, and in particular for the starburst spectra, we find that dust extinction causes a larger shift within the K-L plot than for B-C spectra, probably due to the fact that young or starburst galaxies have a higher gas content; this explains also why the points in that region of the K-L plot display a broader distribution: because of the higher gas content of the galaxies represented in that region, extinction causes larger shifts in the intensity of emission lines and in the slope of the continua.\\

\subsection{Spectral sequence}\label{sec:groupfinding}

\begin{figure}
\includegraphics[scale=1.2]{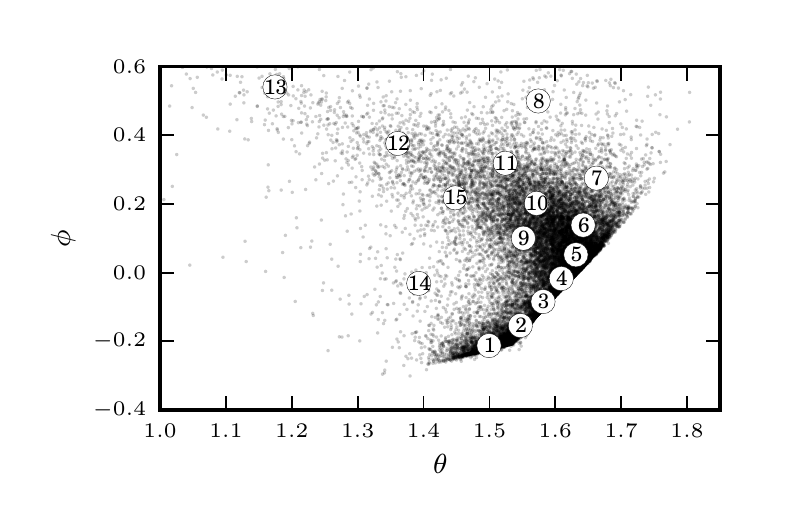}
\caption{\small{K-L plot of VIPERS repaired and cleaned galaxies, labelled with numbers 1-15, that represent the diversity of spectral types. The primary locus is traced by markers 1-8, and we find a secondary branch, marked 9-13. The mean spectrum at each marker is plotted in Fig. 15. }\label{fig:groups}}
\end{figure}

\begin{figure*}
\includegraphics[scale=1.]{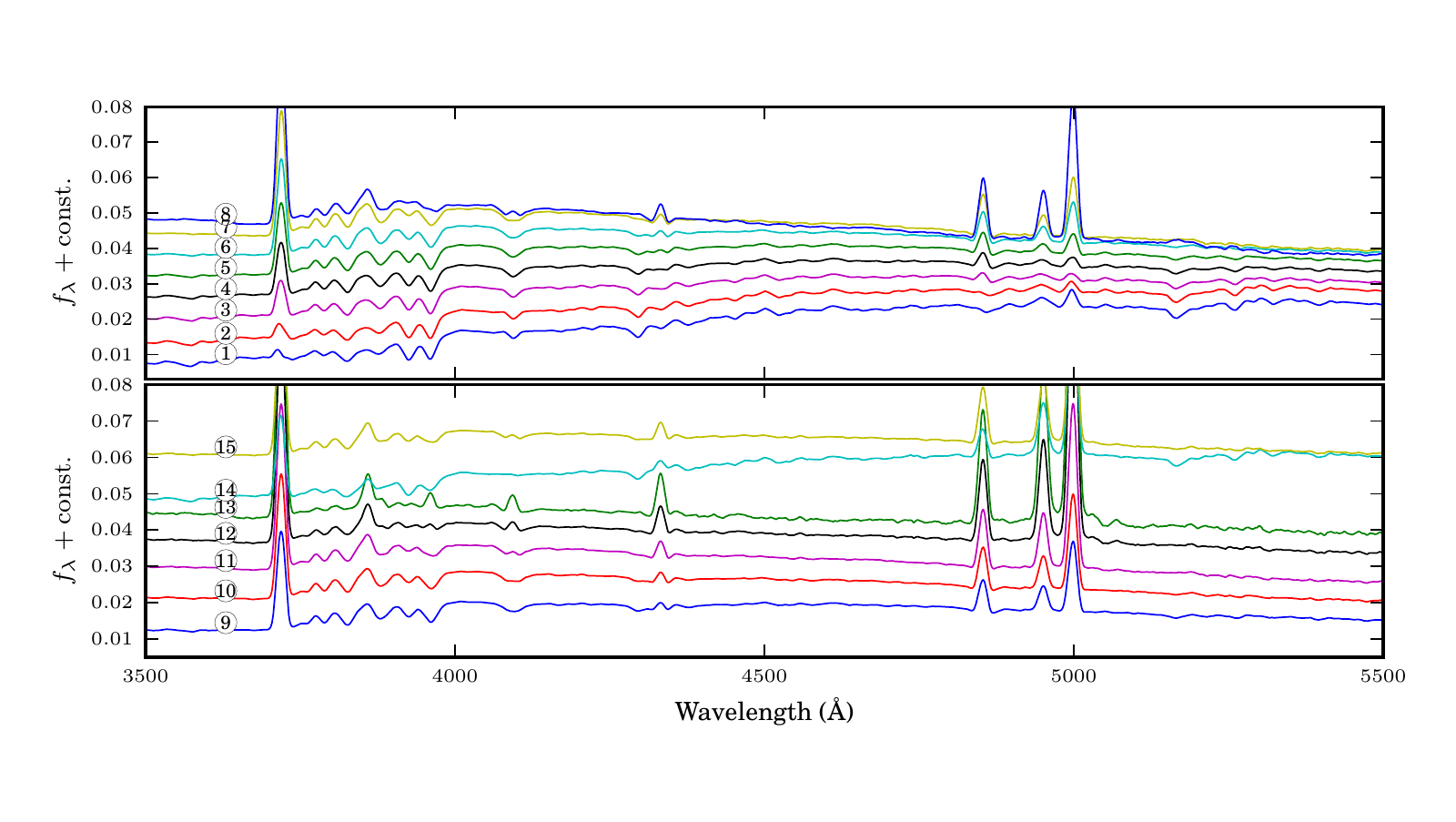}
\caption{\small{Representative average spectra obtained by grouping
    the VIPERS spectra through a group-finding algorithm into 15 classes in the $(\theta, \phi)$ plane, as
    labelled in Fig.~\ref{fig:groups}.  We average the repaired and cleaned
    spectra (i.e. considering only the three principal components).
    In the top frame, we show that spectra 1-8 
    follow a sequence from early to late types, with the continuum becoming
    progressively bluer and with stronger [OII] emission. Note that
    the spectrum labelled as 1, i.e. the reddest one, still presents
    a hint of emission lines (although pure red spectra exist in the
    sample), since it is an average spectrum. In
    the bottom frame, spectra 9-13 represent starburst 
    galaxies with flatter continua and strong emission lines. Mean
    spectra 14-15 effectively seem to pertain to none of the two
    branches, showing a mixture of blue and red galaxy properties.}\label{fig:sequence}} 
\end{figure*}

To explore the diversity of spectra represented on the K-L plot, we
apply a k-means group-finding algorithm that partitions the space into
maximally diverse classes \citep{Ascasanch}.  Galaxies are associated with
a group based upon the distance in the $\theta-\phi$ coordinates.
It is necessary to
specify the number of groups beforehand, and we chose 15, which appear
to be sufficient to span all features visible by eye.

The positions of the classes we have identified are marked in
Fig. \ref{fig:groups}.  These points trace out essentially two
branches that can be thought of as the skeleton of the data cloud. 
The first branch, marked by the numbers 1-8, shows a sequence very
similar to what we can imagine as the prosecution of the B-C red and
intermediate models discussed previously, encompassing though also the
starburst galaxies type 3-4-5-6. In particular, the Sc template
appears to lie between the 7 and 8 classes. A second branch, marked by
9-13, lies almost perpendicular and passes through the starburst 1-2.
The mean spectrum that represents each class is plotted in
Fig. \ref{fig:sequence}. In particular, in the top panel of Fig. \ref{fig:sequence}, we see that moving from 1 to 8 means an increase in the intensity of emission lines and a change in the slope of the continuum, from redder to bluer. In the bottom panel of Fig. \ref{fig:sequence}, mean spectra from 9 to 13, pertaining to the perpendicular ``starburst" branch, show an increase in the intensity of emission lines, particularly evident by looking at the H$\gamma$ emission, while the slope of the continuum is substantially unchanged. 

Consecutive numbers here label very similar average spectra in almost all cases, apart from spectra 14-15, which do not resemble spectrum 13. Mean spectra 14-15 in fact, lying beyond the imaginary starburst branch in Fig. \ref{fig:groups}, actually don't follow the trend of that branch, but show redder continua, in agreement with their $\phi$ position on the K-L plot. They look more similar to mean spectra 3 and 7 respectively, but for the intensity of emission lines, since they exhibit stronger line features. The combination of red continua and strong emission lines shown by mean spectra 14-15, makes them hardly includable in any of the two branches. This suggest that, while moving upwards in the $\phi$ direction in the K-L plot can be associated to a change in the slope and the intensity of the lines, moving from right to left in the $\theta$ direction also means a strengthening in the intensity of the emission lines. 

The shape of the mean spectra for the different groups and the
position of the same groups on the K-L plot reinforce the evidence
that galaxies can be to split into two nearly
orthogonal spectral sequences, of which one reflects the evolutive
phases of a normal galaxy (though not being an evolutionary track),
while the other describes the starburst phases.  This 
suggests a route for building a physical classification of
the spectra based on the K-L parameters, which we plan to develop in a
future work.

\subsection{Comparison with photometric classification}
\begin{figure}
\includegraphics[scale=1.2]{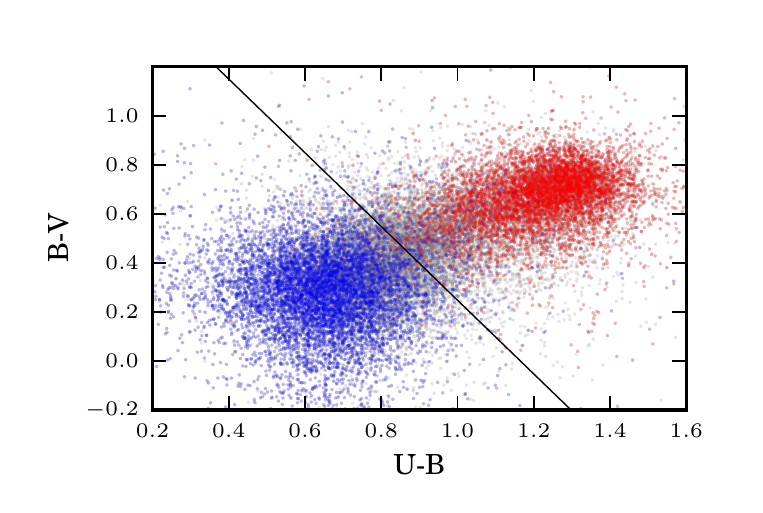}
\caption{\small{The rest frame $U-B$, $B-V$ colours of VIPERS
    galaxies.  Red points have PCA parameter $\phi<-0.1$ and blue
    points have $\phi>0.1$ (intermediate values of $\phi$ are coloured
    grey).  The line dividing the two samples optimally separates
    $\phi>0$ from $\phi<0$ in colour space with a contamination of
    $\sim13$\%.}\label{fig:phot}}
\end{figure}

Finally, we compare side by side the PCA classification against the more familiar one based on rest-frame broad-band photometric colours.
In Fig. \ref{fig:phot} we plot the VIPERS rest-frame $U-B$ and $B-V$
for each galaxy (Bolzonella \emph{et al.}, in prep, Fritz \emph{et
  al.}, in prep.). We divide the sample into red and blue classes
using the K-L angle $\phi$. Based on the comparison to the model
spectra and the discussion of the previous section, a reasonable
definition of the red class can be $\phi<-0.1$, with the blue galaxies
confined at $\phi>0.1$. In this way, we cleanly exclude intermediate
types. 

For comparison, we construct a red-blue classification using the $U-B$
and $B-V$ colours that match as well as possible the PCA
selection. This is shown in Fig. \ref{fig:phot}, where the two classes
defined through the K-L angle are plotted in blue and red and the
intermediate types in grey.  We clearly note that the PCA selection is
correctly capturing the bimodal distribution.  Conversely, let us
verify how a crude color-color selection performs, with respect to
that based on the spectral information ``compressed'' into the PCA
parameters.  We therefore separate photometrically red and blue
classes by tracing a line
perpendicular to the axis connecting the centres of the two clouds,
(Fig. \ref{fig:phot}).  This axis is defined 
by computing, through a simple PCA, the two eigenvectors of the
distribution of points on the colour plane: the first eigenvector
marks the principal direction of the data, while the second is
orthogonal to the first one. Here the total number of eigenvector is
only two, since the correlation matrix of a two-dimensional
distribution has dimension 2. 
The position of the line is set such that there is an approximately
equal number of contaminating galaxies on the red and blue sides. 
With respect to the PCA classification, we find that: (1) in selecting red
galaxies, the color-color selection has a $\sim 14$ per cent
contamination of spetroscopically blue galaxies and an $\sim 88$ per cent completeness;
(2) for photometrically blue galaxies, the contamination of objects
that spectroscopically are classified as ``red'' is $\sim12$ per cent and the
completeness is $\sim 86$ per cent.  

It is encouraging that in this simple case of classifying galaxies as
red or blue, the two methods produce very similar results.  The
strength of the PCA approach is that it encodes additional information
about spectral features that is not available in the broad band
photometry.

\subsection{Outliers} \label{sec:agn}
\begin{figure}
\includegraphics[scale=1.2]{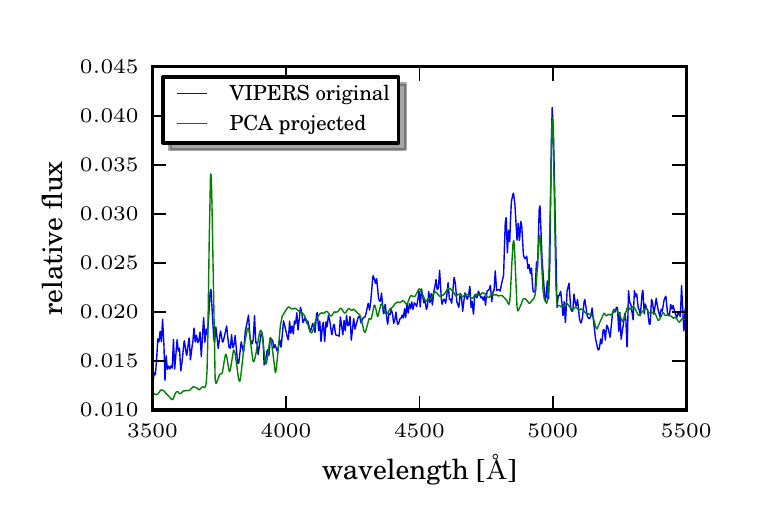}
\caption{\small{Example of an AGN in the VIPERS sample (blue)
    projected on to the PCA eigenspectra basis (green). The PCA
    reconstruction was not able to preserve the peculiarities of this
    rare spectrum, forcing it to resemble a typology of galaxy which
    is much more common within the VIPERS sample.}}\label{fig:agn} 
\end{figure}
One of the limitations of the PCA reconstruction of spectra is that a
spectral type that is represented by a few galaxies only will be
poorly (or even will be not) represented by principal
eigenspectra. Rare features will not be included in the main
eigenspectra, but only in higher-order ones. 
 
This is for example the case of AGNs (as QSOs or
Seyfert galaxies; it can be also the case of normal galaxies which have been assigned a wrong redshift). Their representation, in terms of the first
three components only, will not be realistic. This will force them to
resemble an intermediate, blue, or starburst galaxy.  An example is
shown in 
Fig. \ref{fig:agn}, where a  broad-line AGN is reconstructed using
only three eigenspectra.  The continuum is approximately fitted, but
the broad emission features do not have counterparts in the three
basis vectors used.    

We have directly verified that AGN features start to
emerge only when principal components up to orders $\gtrsim$ 50 are
included. 
This is due to the fact that
AGNs are actually a minority in the VIPERS catalogue (we expect them to be $\sim 5$ per cent of the total), so their peculiar
features are treated as ``noise'' (i.e. uncommon features) by the PCA. For these reasons
the AGNs do not group as a separate population of outliers in the K-L
plot computed with three or higher-order eigenspectra, but fall on the
main locus in apparently random positions.  A PCA reconstruction of
the AGN spectra will be better performed when a 
larger sample of AGNs only will be available.  On the other hand,
given a large data set like VIPERS, for the same reasons the PCA
allows us to identify rare  objects (as the AGN in this case) or even
to look for previously unknown types.  

One could use the goodness of fit $\chi^2$ value to
isolate spectra that are poorly represented by the principal
eigenspectra.  When the $\chi^2$ is larger than a given threshold,
we know that the original is poorly traced by the projection.
A large $\chi^2$ depends also on the signal-to-noise ratio
of the original spectrum. Thus, isolating the spectra presenting a
high $\chi^2$ together with a reasonably high signal-to-noise (as
defined in Eq. \ref{eq: sn}) will select highly-confident outliers.
It will be interesting to explore the application of this technique to
a future, larger version of the VIPERS catalogue and compare it to
alternative methods. 

\section{Summary and Conclusions}
We have developed an objective spectral classification system based on
a principal component analysis for the ongoing VIPERS survey.  Here,
we present the analysis of the first sub-set consisting of 27,350
galaxy spectra at redshifts $0.4 < z < 1.0$.  Our implementation
of a principal component analysis addresses the non-uniform
characterstics of the dataset that can impede the measurement and
classification of spectral features, including the variation of
wavelength coverage in the rest frame, noise properties and
instrumental artefacts. We correct for these effects using an
iterative algorithm that converges to a robust estimate of the
eigenspectra templates.

Our final classification system is based upon three coefficients,
$a_1$, $a_2$ and $a_3$, that are found by projecting the spectra on to
the first three principal components. The determination of the
coefficients for each spectrum uses a specific recipe to preserve the
physicality of spectral lines such that both the continuum and line
features are reconstructed accurately.  The first three
eigencoefficients provide a high-fidelity reconstruction of the
spectrum for a broad range of galaxy types.

The information enclosed in the three eigencoefficients can be
compressed in the K-L angles representation: $\phi=\tan^{-1}(a_2/a_1)$
and $\theta=\cos^{-1}a_3$. This is a key step for our spectral
classification: in a $\theta$--$\phi$ plane galaxies of different
colour concentrate in different regions, according to the relative
importance of the three eigenspectra. These, at least in terms of the
continuum, mirror the shape of realistic red, blue and intermediate
galaxies.

To explore the physical meaning of the different positions on the
$\theta$--$\phi$ diagram, we projected a set of Bruzual-Charlot model spectra on the
same VIPERS eigenspectra and looked at their distribution on the same
plot. We also added a set of 12 Kinney-Calzetti templates, as to
verify the appearance of starburst galaxies over the same plane.  An
analysis with a group finding algorithm capable to 
divide space into maximally diverse classes, showed clear
evidence of two different branches, following respectively the trend
of the Bruzual-Charlot and Kinney-Calzetti models. The models have
been also dust extincted to know in which direction the 
reddening for spectra moves the points in the K-L cloud. 

A comparison of our classification method with a more common
photometric selection shows that the PCA approach is
comparable to a rest-frame color-color plot in discriminating red from blue
galaxies, whereas being more sensitive than photometry to intermediate
spectral types, being based on spectra. 

Some peculiar spectra will not be well represented in the
eigenspectra, due to the rareness of their features in the sample.
For instance, we find that the eigenspectra do not fit AGN spectra
well.  However, in principle, interesting outlying spectra can be
identified based upon poor $\chi^2$ values for the fit.

We remark that we have analysed only the initial 40 per cent of the VIPERS
survey. As the data sample increases and the statistics grow, the
repairing procedure will improve in precision.  In future analyses we
will have the possibility to divide the sample in redshift bins.
Additionally, the analysis can be naturally extended to include
additional observations, such as galaxy luminosities and broadband
fluxes.

\section*{Acknowledgments}

We acknowledge the support of the ESO staff to VIPERS through
service-mode observing, in particular our support astronomer, M.
Hilker. 

We acknowledge financial support through grants PRIN INAF 2008
and PRIN INAF 2010. KM, AP, JK have been supported by the research grant of
the Polish Ministry of Science N N203 512938. A part of this work was
carried out within the framework of the European Associated Laboratory
"Astrophysics Poland-France". AP has been partially supported 
by the project POLSIH- SWISS ASTRO PROJECT co-financed by a grant from
Switzerland through the Swiss Contribution to the enlarged
European Union. KM has been supported from the Japan
Society for the Promotion of Science (JSPS) Postdoctoral Fellowship
for Foreign Researchers, P11802. GDL acknowledges financial support from the European Research Council under
the European Community's Seventh Framework Programme (FP7/2007-2013)/ERC grant
agreement n. 202781.

Based on observations obtained with MegaPrime/MegaCam, a joint project 
of CFHT and CEA/DAPNIA, at the Canada-France-Hawaii Telescope (CFHT) 
which is operated by the National Research Council (NRC) of Canada, the 
Institut National des Science de l'Univers of the Centre National de la 
Recherche Scientifique (CNRS) of France, and the University of Hawaii. 
This work is based in part on data products produced at TERAPIX and the 
Canadian Astronomy Data Centre as part of the Canada-France-Hawaii 
Telescope Legacy Survey, a collaborative project of NRC and CNRS.

\setlength{\bibhang}{2.0em}
\setlength\labelwidth{0.0em}
\bibliographystyle{mn2e}
\footnotesize{

}

\label{lastpage}

\end{document}

%% file: authors_gigi_work.tex
\author[A. Marchetti et al.]{A. 
Marchetti$^{1,2}$\thanks{E-mail: \href{mailto:alida.marchetti@brera.inaf.it}{alida.marchetti@brera.inaf.it}},
B. R. Granett$^{2}$,
  L. Guzzo$^{2}$,
  A. Fritz$^{3}$,
  B. Garilli$^{3,4}$,
  M. Scodeggio$^{3}$,
  \newauthor
U. Abbas$^{5}$,                 
  C. Adami$^{4}$,               
  S. Arnouts$^{6}$,             
M. Bolzonella$^{9}$,           
 D. Bottini$^{3}$,             
 A. Cappi$^{9}$,
 J. Coupon$^{12}$,                              
\newauthor
O. Cucciati$^{13}$,           
 G. De Lucia$^{13}$,
  S. de la Torre$^{14}$,
P. Franzetti$^{3}$,
 M. Fumana$^{3}$,
O. Ilbert$^{4}$,
\newauthor
 A. Iovino$^{2}$,
  J. Krywult$^{15}$,
  V. Le Brun$^{4}$,
  O. Le Fevre$^{4}$,
D. Maccagni$^{3}$,
 K. Malek$^{16,21}$,
 \newauthor
 F. Marulli$^{9,17,18}$,
 H. J. McCracken$^{19}$,
  B. Meneux$^{20}$,
L. Paioro$^{3}$,
 M. Polletta$^{3}$,
  \newauthor
 A. Pollo$^{22,23}$,
 H. Schlagenhaufer$^{24,20}$,
 L. Tasca$^{4}$,
 R. Tojeiro$^{11}$,
  D. Vergani$^{25,9}$,
 \newauthor
A. Zanichelli$^{26}$,
  J. Bel$^{7}$,
  M. Bersanelli$^{1}$,
  J. Blaizot$^{8}$,
E. Branchini$^{10}$,
  A. Burden$^{11}$,
\newauthor
 I. Davidzon$^{9,17,18}$,
 C. Di Porto$^{9}$,
 L. Guennou$^{4}$,
  C. Marinoni$^{7}$,
Y. Mellier$^{19}$,
 \newauthor
L. Moscardini$^{9,17,18}$,
 R.C. Nichol$^{11}$,
  J.A. Peacock$^{14}$,
 W.J. Percival$^{11}$,
  S. Phleps$^{20}$,
\newauthor
C. Schimd$^{4}$,
  M. Wolk$^{19}$,
G. Zamorani$^{9}$
\\
\\
    $^{1}$ Universit\'{a} degli Studi di Milano, via G.Celoria 16, 20130 Milano, Italy\\
    $^{2}$ Istituto Nazionale di Astrofisica - Osservatorio Astronomico di Brera, Via Brera 28, 20122 Milano, Via E. Bianchi 46, 23807 Merate, Italy\\
   $^{3}$ Istituto Nazionale di Astrofisica - Istituto di Astrofisica Spaziale e Fisica Cosmica Milano, via Bassini 15, 20133 Milano, Italy\\
$^{4}$ Aix Marseille Universit\'{e}, CNRS, LAM (Laboratoire
    d'Astrophysique de Marseille) UMR 7326, 13388, Marseille, France\\ 
    $^{5}$ INAF - Osservatorio Astronomico di Torino, 10025, Pino Torinese, Italy\\
    $^{6}$ Canada-France-Hawaii Telescope, 65--1238 Mamalahoa Highway, Kamuela, HI 96743, USA\\
    $^{7}$ Centre de Physique Th\'eorique, UMR 6207 CNRS-Universit\'e de Provence, Case 907, F-13288 Marseille, France\\
    $^{8}$ Universit\'{e} de Lyon, Lyon F-69003, France\\
    $^{9}$ Istituto Nazionale di Astrofisica - Osservatorio Astronomico di Bologna, via Ranzani 1, I-40127, Bologna, Italy\\
    $^{10}$ Dipartimento di Fisica 'E. Amaldi', Universit\'{a} degli Studi 'Roma Tre', via della Vasca Navale 84, 00146 Roma, Italy\\
    $^{11}$ Institute of Cosmology and Gravitation, Dennis Sciama Building, University of Portsmouth, Burnaby Road, Portsmouth, PO1 3FX\\
    $^{12}$ Institute of Astronomy and Astrophysics, Academia Sinica, P.O. Box 23-141, Taipei 10617, Taiwan\\
    $^{13}$ Istituto Nazionale di Astrofisica - Osservatorio Astronomico di Trieste, via G. B. Tiepolo 11, 34143 Trieste, Italy\\
    $^{14}$ SUPA, Institute for Astronomy, University of Edinburgh, Royal Observatory, Blackford Hill, Edinburgh EH9 3HJ, UK\\
    $^{15}$ Institute of Physics, Jan Kochanowski University, ul. Swietokrzyska 15, 25-406 Kielce, Poland \\
    $^{16}$ Department of Particle and Astrophysical Science, Nagoya University, Furo-cho, Chikusa-ku, 464-8602 Nagoya, Japan\\
    $^{17}$ Dipartimento di Astronomia, Alma Mater Studiorum - Universit\`{a} di Bologna, via Ranzani 1, I-40127 Bologna, Italy\\
    $^{18}$ INFN/National Institute for Nuclear Physics, Sezione di Bologna, viale Berti Pichat 6/2, I-40127 Bologna, Italy\\
    $^{19}$ Institute d'astrophysic de Paris, UMR7095 CNRS, Universit\`{e} Pierre et Marie Curie, 98 bis Boulevard Arago, 75014 Paris, France\\
    $^{20}$  Max-Planck-Institut f\"{u}r Extraterrestrische Physik, D-84571 Garching b. M\"{u}nchen, Germany\\
    $^{21}$ Center for Theoretical Physics of the Polish Academy of Sciences, Al. Lotnikow 32/46, 02-668 Warsaw, Poland\\
    $^{22}$ Astronomical Observatory of the Jagiellonian University, Orla 171, 30-001 Cracow, Poland\\
    $^{23}$ National Centre for Nuclear Research, ul. Hoza 69, 00-681 Warszawa, Poland \\
    $^{24}$ Universit\"{a}tssternwarte M\"{u}nchen, Ludwig-Maximillians Universit\"{a}t, Scheinerstr. 1, D-81679 M\"{u}nchen, Germany\\
    $^{25}$  Istituto Nazionale di Astrofisica Istituto di Astrofsica Spaziale e Fisica Cosmica Bologna, via Gobetti 101,I-40129 Bologna, Italy\\
    $^{26}$ Istituto di Radioastronomia Istituto Nazionale di Astrofisica, via Gobetti 101, I-40129, Bologna, Italy\\
}